\documentclass{PoS}

\usepackage{epsfig}
\usepackage{latexsym}
\usepackage{amsmath}
\usepackage{amsfonts}
\title{Renormalization constants for Lattice QCD: new results from Numerical Stochastic Perturbation Theory}

\ShortTitle{Renormalization constants for Lattice QCD: new results from Numerical Stochastic Perturbation Theory}

\author{\speaker{Francesco Di Renzo}\\
        University of Parma and I.N.F.N.,
        Viale Usberti 7/A, I-43100 Parma, Italy\\
        E-mail: \email{direnzo@fis.unipr.it}}
\author{Vincenzo Miccio\\
        Sezione I.N.F.N. Milano Bicocca,
        Piazza della Scienza 3, I-20126 Milano, Italy\\
        E-mail: \email{vincenzo.miccio@mib.infn.it}}
\author{Luigi Scorzato\\
        ECT* and I.N.F.N.,
        Strada delle Tabarelle 286, I-38050 Villazzano (Trento), Italy\\
        E-mail: \email{scorzato@ect.it}}
\author{Christian Torrero\\
        Faculty of Physics, University of Bielefeld, D-33501 Bielefeld, Germany\\
        E-mail: \email{torrero@physik.uni-bielefeld.de}}

\abstract{By making use of Numerical Stochastic Perturbation Theory (NSPT) we can compute renormalization
constants for Lattice QCD to high orders, \emph{e.g.} three or four loops for quark bilinears. We report on 
the status of our computations, which provide several results for Wilson quarks and in particular (values and/or 
ratios of) $Z_V$, $Z_A$, $Z_S$, $Z_P$. Results are given for various number of flavors ($n_f = 0, 2, 3, 4$). While we recall the care which is due for the computation of quantities for which an anomalous dimension is in place, we point out that our computational framework is well suited to a variety of other calculations and we briefly discuss the application of NSPT to other regularizations (in particular the Clover action).}

\FullConference{XXIVth International Symposium on Lattice Field Theory\\
                July 23-28, 2006\\
                Tucson, Arizona, USA}

\begin{document}

\section{Introduction}
Numerical Stochastic Perturbation Theory (NSPT) is a numerical tool which enables us to perform high order 
computations in Lattice Perturbation Theory (for a comprehensive introduction see \cite{NSPTfull}). In recent 
years our group undertook the high order computation of Lattice QCD renormalization constants. 
We have till now mainly focused on Wilson quarks bilinears \cite{Lat05,cyprus}, a task which is by now 
almost completed (\cite{ourZeds} will contain a comprehensive report). Here we report on the three (or even four) loop calculation of finite renormalization constants. The overall strategy of the computation will be sketched in order to 
make this communication sufficiently self consistent.

\section{Computational setup}
Our computations are for Wilson gauge action and (unimproved) Wilson fermions. 
We can rely on a collection of NSPT configurations for various number of flavors: $n_f=0$ (three loops, both on $32^4$ and on $16^4$ lattices; the latter is mainly used to check finite size effects); $n_f=2$ (on a $32^4$ lattice, to three and even four loops); $n_f=3$ and $n_f=4$ have at the moment only a reduced sample of configurations, which enables us to quote preliminary results (\emph{i.e.} only for certain quantities).

The scheme we adhere to is what is known as \emph{RI'-MOM}, which became very popular after its non-perturbative implementation \cite{RI-MOM-rome}. It is important to point out that a comprehensive computation of the quark bilinears anomalous dimensions for this scheme is available in the literature \cite{JGrac}.

In order to implement our program we start from the computation of quark bilinears between external quark states at fixed (off-shell) momentum $p$
\begin{equation}
\int dx \,\langle p | \; \overline{\psi}(x) \Gamma \psi(x) \; | p \rangle \, = \, G_{\Gamma}(pa), 
\end{equation}
where $\Gamma$ stands for any of the $16$ Dirac matrices singling out the {$S,V,P,A,T$} currents. We notice that 
on the lattice the dependence is on $pa$.
These quantities are gauge-dependent and our computations are in the Landau gauge, which is easy to fix on the lattice and does not require to discuss the gauge parameter renormalization. 
Next step is trading the $G_{\Gamma}(pa)$ for the amputated function, in our notation $\Gamma_{\Gamma}(pa)$
\begin{equation}
G_{\Gamma}(pa) \;\;
\rightarrow \;\; \Gamma_{\Gamma}(pa) \, = \, S^{-1}(pa) \; G_{\Gamma}(pa) \; S^{-1}(pa).
\end{equation}
The $\Gamma_{\Gamma}(pa)$ are finally projected on the tree-level structure by a suitable operator $\hat{P}_{O_{\Gamma}}$
\begin{equation}
	O_{\Gamma}(pa) = \mbox{Tr}\left(\hat{P}_{O_{\Gamma}} \; \Gamma_{\Gamma}(pa)\right).
\end{equation}
Renormalization conditions are given in terms of the $O_{\Gamma}(pa)$ according to the master formula
\begin{equation}\label{master}
	Z_{O_{\Gamma}}(\mu a, g(a)) \, \; Z_q^{-1}(\mu a, g(a)) \, \; O_{\Gamma}(pa) \Big|_{p^2 = \mu^2} \, = \, 1.	
\end{equation}
The dependence of the $Z$'s on the scale $\mu$ is via the dimensionless quantity $\mu a$, while the dependence on $g(a)$ explicitly displays that our results will be expansions in the lattice coupling. The quark field renormalization constant ($Z_q$) entering the above formula is defined by
\begin{equation}\label{Zq}
	Z_q(\mu a, g(a)) \, = \, - i \frac{1}{12} \frac{\mbox{Tr}(\hspace{-.2em}\not\hspace{-.2em}p \; S^{-1}(pa))}{p^2}\Big|_{p^2 = \mu^2}.
\end{equation}
In order to get a mass-independent renormalization scheme, one gives the renormalization conditions in the massless limit. In Perturbation Theory this implies the knowledge of the relevant counterterms, \emph{i.e.} the Wilson fermions critical mass. One and two loops results are necessary in order to implement our three loops computations and are known from the literature \cite{PanaPelo}. As a side remark we notice that third (and fourth) loops have been computed by us as a (necessary) byproduct of the current computations. We stress that in our (perturbative) framework reaching the massless limit does not require any extrapolation procedure.

At any loop order $L$, the expected form of the $L^{th}$ coefficient of a renormalization constant is 
\begin{equation} \label{CC}
	z_L = c_L + \sum_{i}^{L} d_i(\gamma) \log(\hat{p})^i + F(\hat{p}) \;\;\;\;\;\;\;
\left(\hat{p} = p a \right),
\end{equation}
\emph{i.e.} we have to look for a finite number, a divergent part which is a function of anomalous dimensions $\gamma$'s and irrelevant pieces, which we expect to be compliant to hypercubic symmetry.
The anomalous dimensions which are needed are to be taken from the literature and subtracted. After such a subtraction a \emph{hypercubic symmetric Taylor expansion} enables us to fit the irrelevant pieces given by  $F(\hat{p})$. It is instructive to write down an example: for the scalar current our master formula Eq.~(\ref{master}) reads (in Landau gauge the quark field has zero anomalous dimension at one loop)
\begin{equation}
	z_s^{(1)} = z_q^{(1)} - \left( o_s^{(1)} - \gamma_s^{(1)} 
\log(\hat{p}^2)  \right).
\end{equation}
$o_s^{(1)}$ is the quantity actually measured from our simulations; after subtracting the $\log$ one should now proceed to fit the irrelevant pieces in order to get the constant $z_s^{(1)}$.
This subtraction procedure, as we now briefly comment, is actually affected by finite size effects which are to be assessed. 

It is very important to stress that \emph{RI'-MOM} is an infinite volume scheme, while we are necessarily computing on finite volumes. This results in a \emph{taming} of the $\log$, due to $p L$ effects (notice that $pL = \hat{p}N$, where $N$ is the number of lattice sites). This has been actually verified by comparing $16^4$ and $32^4$ results. The correct one loop result is recovered once one computes the finite size corrections to the $\log$ on a finite volume in the continuum: for a preliminary discussion the interested reader is referred to \cite{cyprus}.

\section{Wilson quark bilinears to three (and even four) loops}

Having pointed out the delicate issue which is in place in the case of anomalous dimensions, we now proceed to discuss the finite cases, for which we do already have definite results. 

\subsection{Ratios are always safe}

Results to three loops for the finite ratios of renormalization constants $Z_P/Z_S$ and $Z_V/Z_A$ are reported in Table 1. 
The quark field renormalization constant present in Eq.~(\ref{master}) cancels out in the ratios, together with the divergences that affect $Z_P$ and $Z_S$ separately. In Fig.~1 we present (one loop) examples of the extrapolation procedure which is entailed in fitting hypercubic symmetric Taylor expansions: powers of $\hat{p}=p a$ are the irrelevant contributions to be discarded in the continuum limit, in which the constant (intercept at $\hat{p}=0$) is singled out. Notice that for $Z_V/Z_A$ one can inspect \emph{families} of curves: for given values of $p^2$ they correspond to different lengths of momentum $p$ in the relevant direction $\mu$ (\emph{e.g.} for $\overline{\psi}\,\gamma^\mu\psi$), \emph{i.e.} they correspond to different irrelevant contributions (for a short discussion see \cite{cyprus}). 

Many checks have been performed on the expansions. Finite size effects are well under control, as checked by comparing results on $16^4$ and $32^4$ lattices. Notice also that one can compute (for example) both $Z_A/Z_V$ and $Z_V/Z_A$. Different ratios come from different (although correlated) combinations of data (this is due to the order by order nature of the computation): to a very good precision the series obtained are inverse of each other. 

%
\begin{figure}[t]
  \begin{center} 
		\includegraphics[scale=0.5]{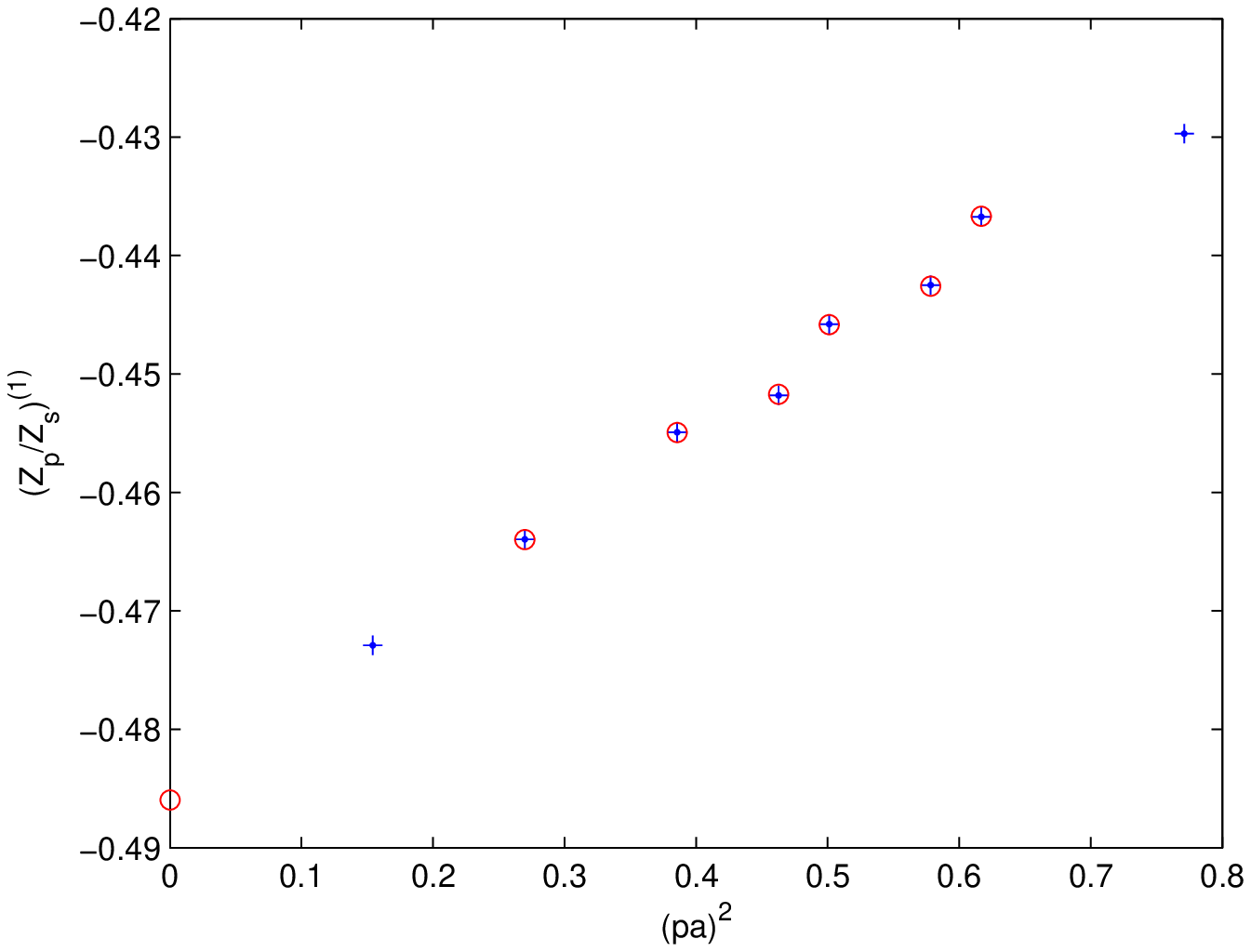}
		\includegraphics[scale=0.5]{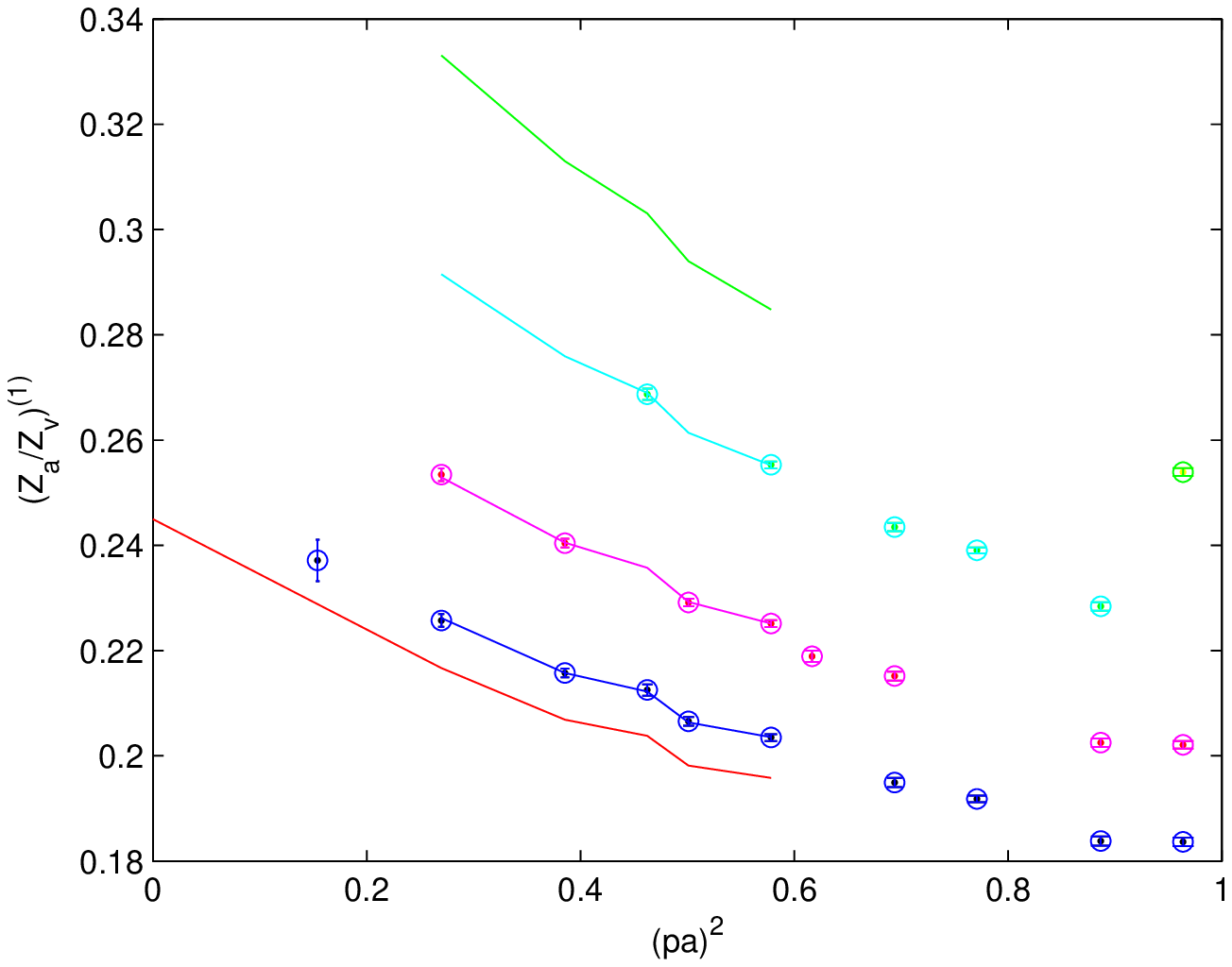}
    \caption{Computation to one loop of finite ratios of renormalization constants: $Z_P/Z_S$ (left) and $Z_A/Z_V$ (right). For the latter different curves correspond to different lengths of the component of momentum $p$ along the relevant direction (see text). Solid lines join the data points which have been taken into account in the fit.}
   \label{Fig.1}
  \end{center}
\end{figure}%

\begin{table}[b]
\begin{center}
\begin{tabular}{|c|c|c|c|}
\hline
$Z_P/Z_S$ & & & \\
\hline
\hline
$n_f$ &  $O(\beta^{-1})$  &   $O(\beta^{-2})$  &   $O(\beta^{-3})$  \\
\hline
0   & - 0.487(1)  & - 1.50(1) & - 5.72(3) \\
\hline
2   &  - 0.487(1)  & - 1.46(1) & - 5.35(3) \\
\hline
3   & - 0.487(1)  & - 1.43(1) & - 5.13(3) \\
\hline
4   &  - 0.487(1)  & - 1.40(1) & - 4.86(3) \\
\hline
\hline
$Z_V/Z_A$  & & &\\
\hline
\hline
$n_f$ &  $O(\beta^{-1})$  &   $O(\beta^{-2})$  &   $O(\beta^{-3})$ \\
\hline
0   & - 0.244(1)  & - 0.780(5) & - 3.02(2) \\
\hline
2   &  - 0.244(1)  & - 0.759(5) & - 2.83(2) \\
\hline
3   & - 0.244(1)  & - 0.744(6) & - 2.72(2) \\
\hline
4   &  - 0.244(1)  & - 0.732(6) & - 2.57(2) \\
\hline
\end{tabular}
\vskip 0.5cm
\caption{The finite ratios of renormalization constants $Z_P/Z_S$ and $Z_V/Z_A$ for various number of flavour $n_f$. The four loops results which are available for $n_f=2$ are not quoted.}
\label{T_PS_VA}
\end{center}
\end{table}

One can also check that the $n_f$ dependence is (within errors) compatible with the expected one (linear at two loop, quadratic at three loop): see Fig.~2. 

To proceed to four loops one needs to plug in the three loops critical mass counterterm. Since we know it to a sufficiently good accuracy for the case $n_f=2$, in this case we also have preliminary four loop results, which we do not quote in this communication.

%
\begin{figure}[t]
  \begin{center} 
		\includegraphics[scale=0.5]{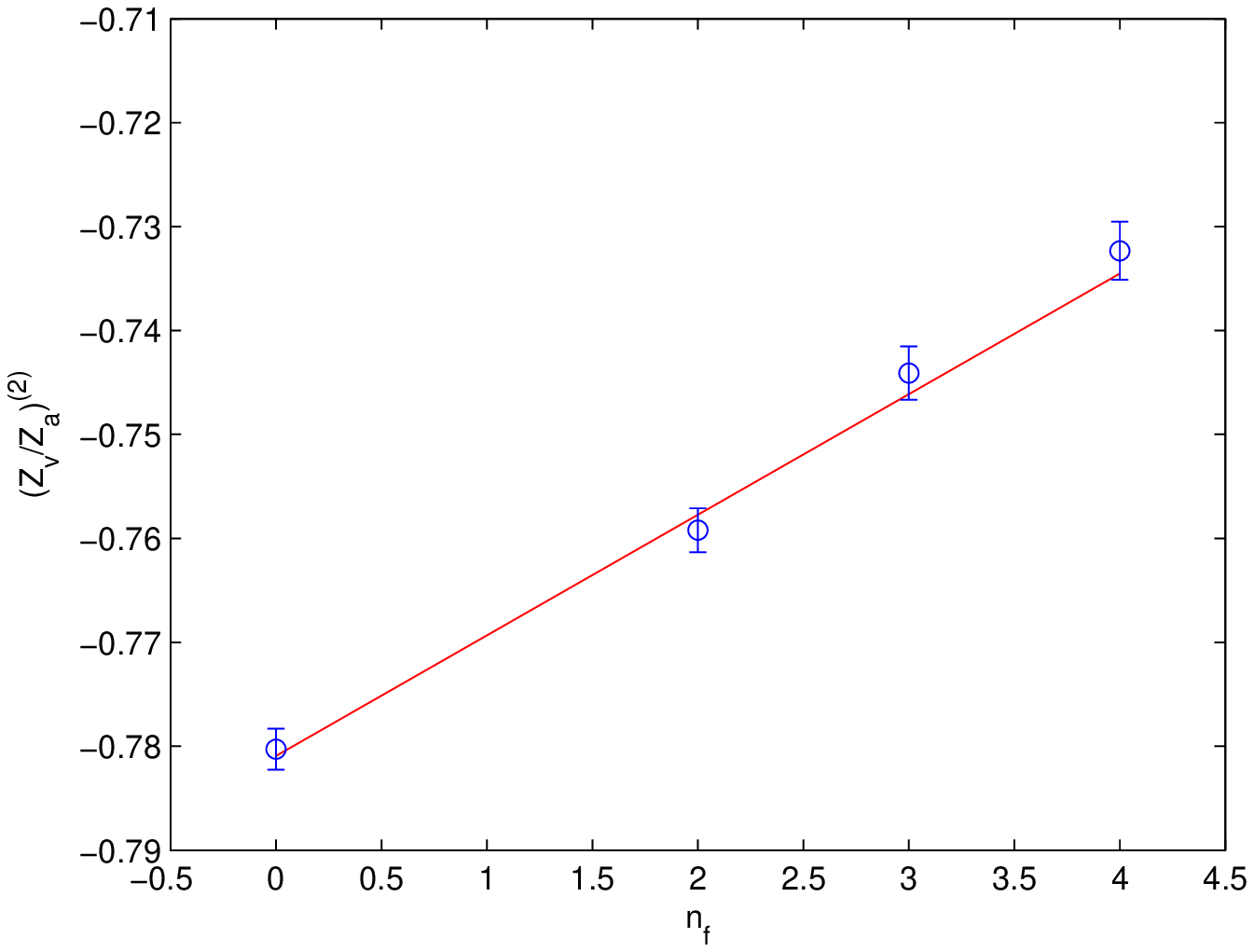}
		\includegraphics[scale=0.5]{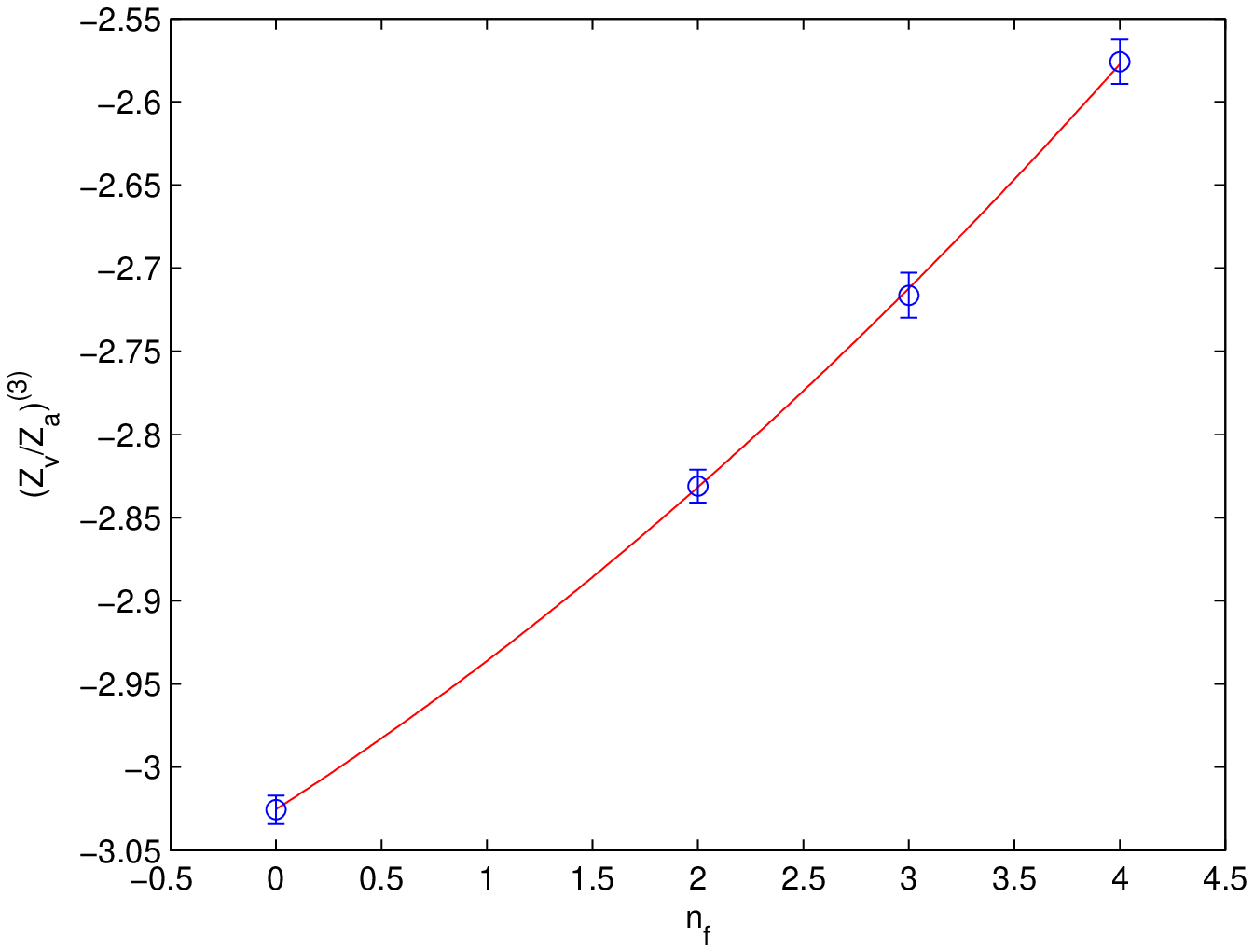}
    \caption{The $n_f$ dependence of the ratio $Z_V/Z_A$ at two (left, linear fit) and three (right, quadratic fit) loops.}
   \label{Fig.2}
  \end{center}
\end{figure}%

\subsection{$Z_V$ and $Z_A$}

One loop examples of computations of $Z_V$ and $Z_A$ are depicted in Fig.~3. $Z_V$ and $Z_A$ are finite by themselves. 
In our master formula Eq.~(\ref{master}) they are interlaced with log's coming from the quark field renormalization constant. The latter can be eliminated either directly from measurements of the propagator or by taking ratios with the conserved vector current. Both procedures return consistent results, which are summarized in Table 2. 

%
\begin{figure}[b]
  \begin{center} 
		\includegraphics[scale=0.5]{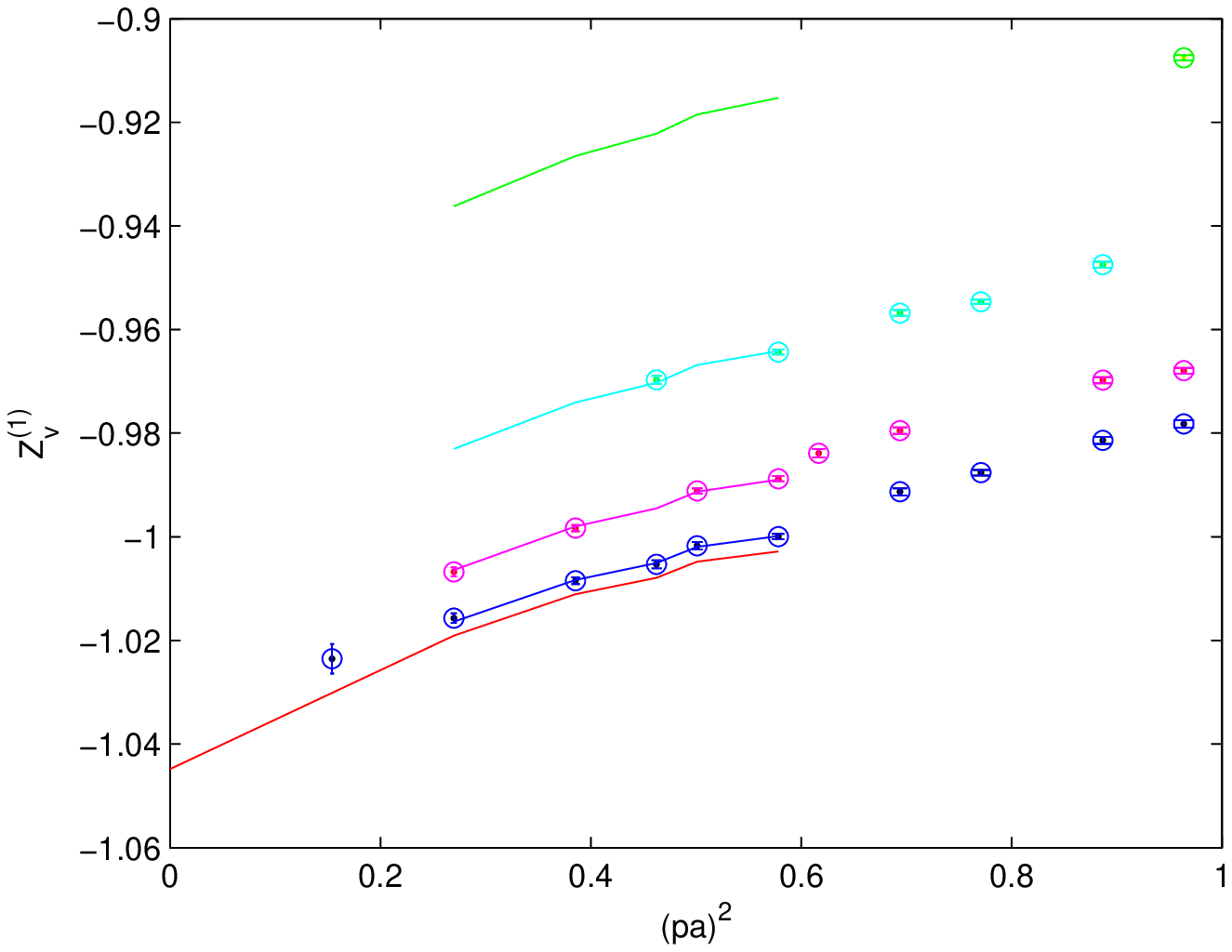}
		\includegraphics[scale=0.5]{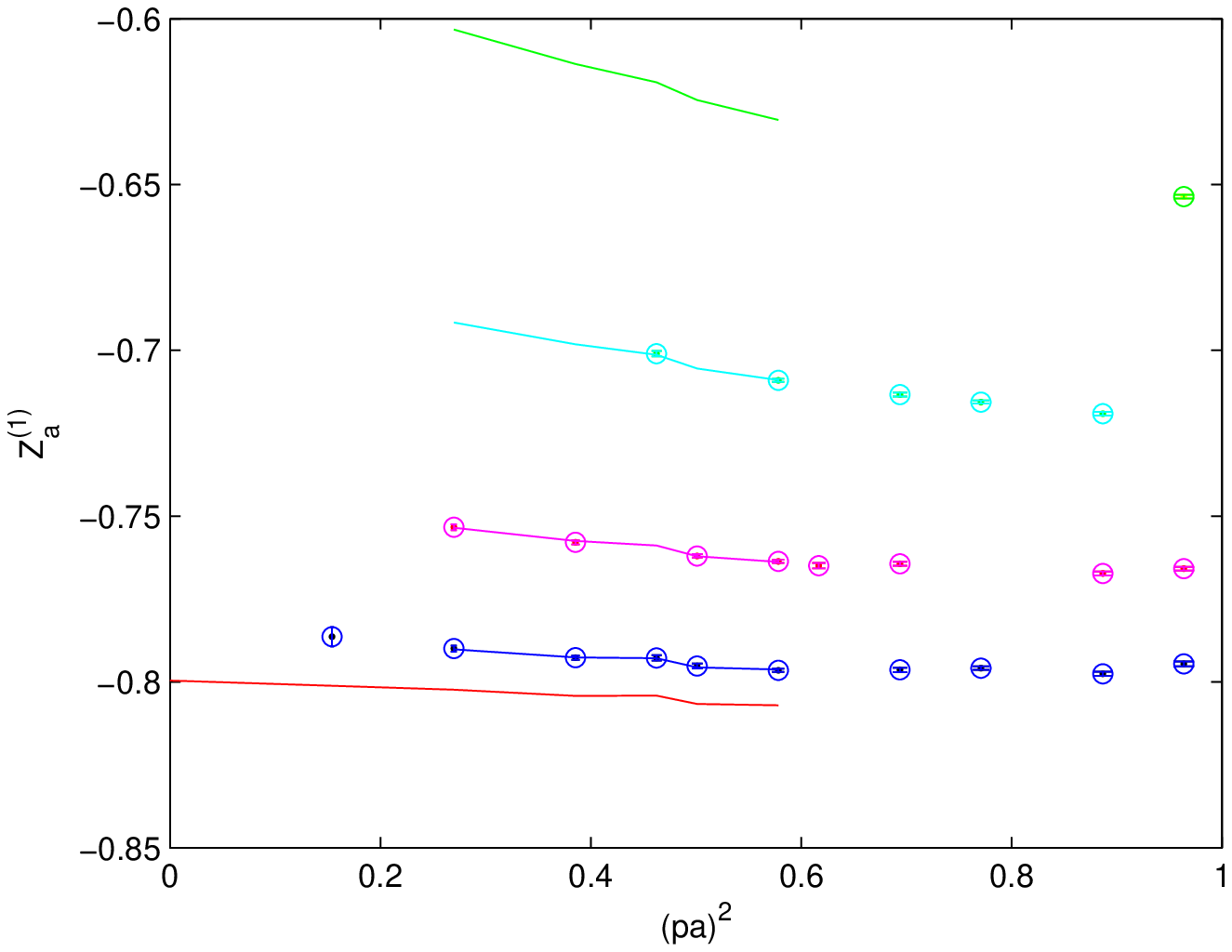}
    \caption{Computation to one loop of finite renormalization constants: $Z_V$ (left) and $Z_A$ (right). As in Fig.~1, different curves correspond to different lengths of the component of momentum $p$ along the relevant direction (see text) and solid lines join the data points which have been taken into account in the fit.}
   \label{Fig.3}
  \end{center}
\end{figure}%

\begin{table}[ht]
\begin{center}
\begin{tabular}{|c|c|c|c|}
\hline
$Z_V$ & & & \\
\hline
\hline
$n_f$ &  $O(\beta^{-1})$  &   $O(\beta^{-2})$  &   $O(\beta^{-3})$ \\
\hline
0   & - 0.487(1)  & - 1.50(1) & - 5.72(3) \\
\hline
2   &  - 0.487(1)  & - 1.46(1) & - 5.35(3)  \\
\hline
\hline
$Z_A$  & & & \\
\hline
\hline
$n_f$ &  $O(\beta^{-1})$  &   $O(\beta^{-2})$  &   $O(\beta^{-3})$  \\
\hline
0   & - 0.244(1)  & - 0.780(5) & - 3.02(2) \\
\hline
2   &  - 0.244(1)  & - 0.759(5) & - 2.83(2) \\
\hline
\end{tabular}
\vskip 0.5cm
\caption{The finite renormalization constants $Z_V$ and $Z_A$ for various number of flavor $n_f$. The four loops results which are available for $n_f=2$ are not quoted.}
\label{T_V_A}
\end{center}
\end{table}

Once again, also 4 loops results are available for $n_f=2$. Fig.~4 displays resummations of $Z_A$ and $Z_V$ at $\beta=5.8$. In this resummations we do take into account also four loops results. In the spirit of Boosted Perturbation Theory (BPT) \cite{BPT}, we show different resummations for different definitions of the coupling ($P$ is the basic plaquette):
\begin{equation}
	x_0 \equiv \beta^{-1} \;\;\;\; x_1 \equiv - \frac{1}{2} \log(P)  \;\;\;\; x_2 \equiv \frac{\beta^{-1}}{P}.
\end{equation}
Our ($n_f=2$, $\beta=5.8$) results are $Z_A = 0.79(1)$ and $Z_V=0.70(1)$ (one can compare to \cite{CeciVitt}). Notice how it is impossible to assess the consistency of a 1 loop BPT result without knowing higher orders in the expansion. 

%
\begin{figure}[b]
  \begin{center} 
		\includegraphics[scale=0.5]{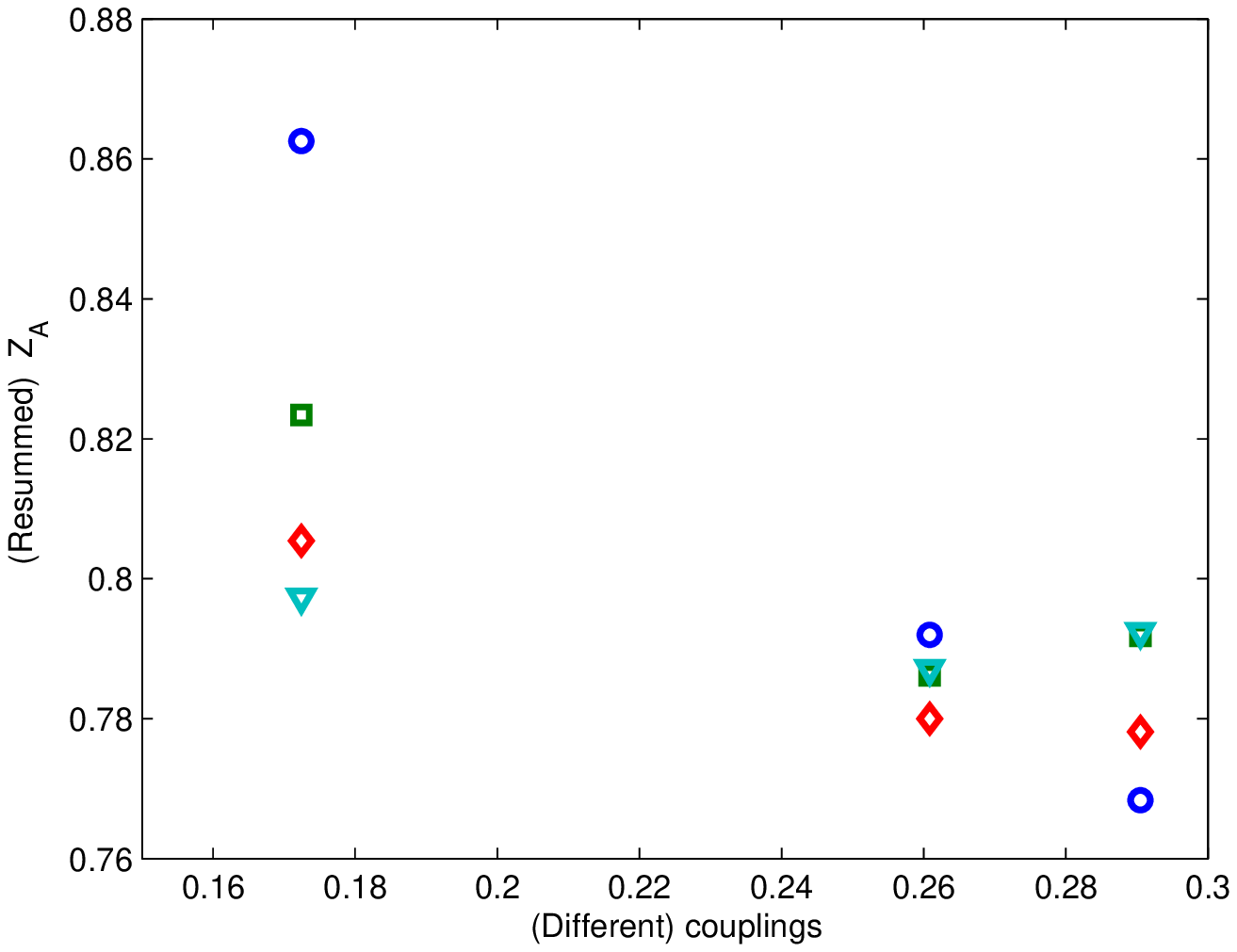}
		\includegraphics[scale=0.5]{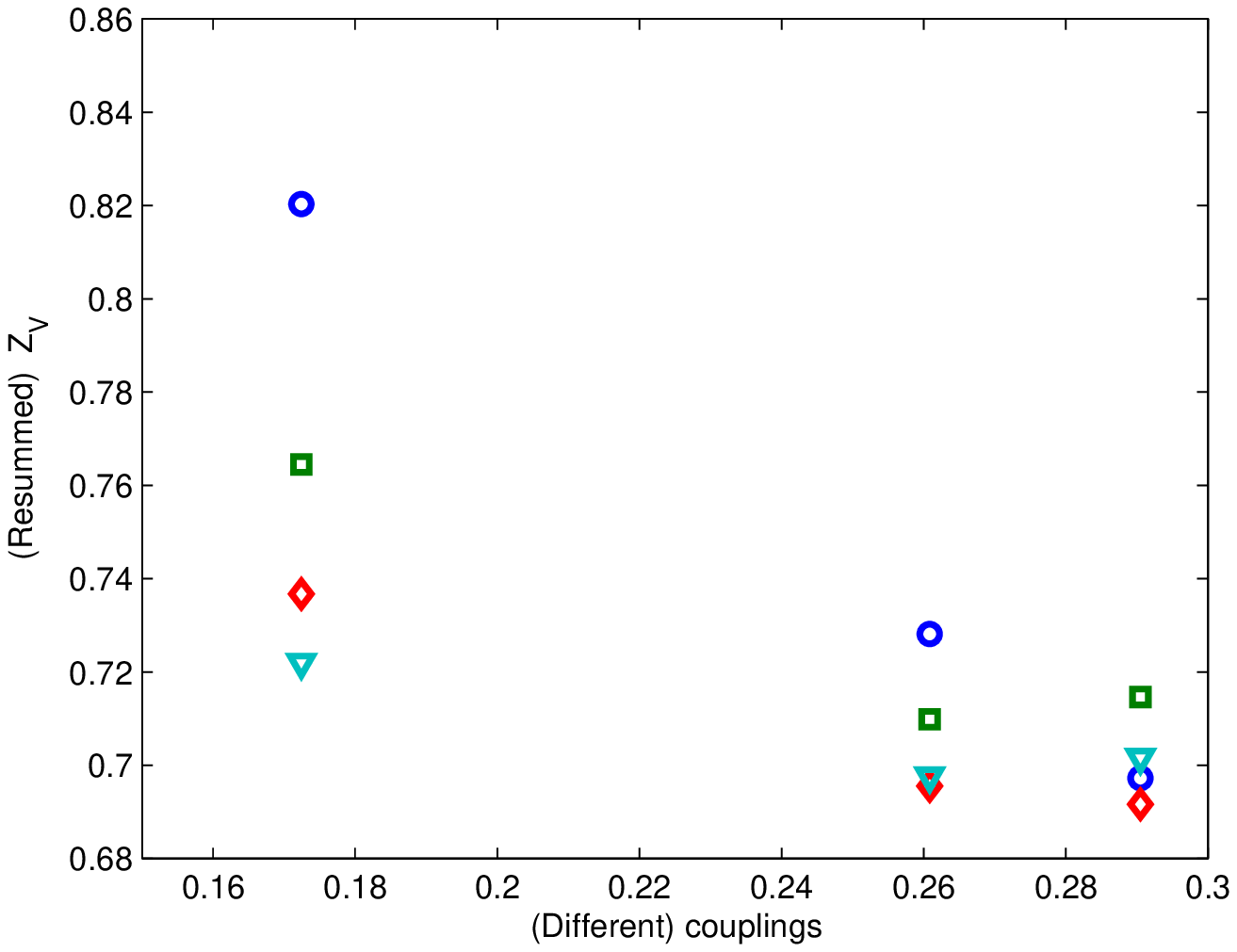}
    \caption{Resummations of $Z_A$ (left) and $Z_V$ (right) for $n_f=2$ at $\beta=5.8$ to one (circles), two (squares), three (diamonds) and four (triangles) loops. We show resummations for different couplings: on the $x$-axis, the (different) values of the different couplings. From the left: $x_0$, $x_1$, $x_2$ ($x_0$ is $\beta^{-1}$, see text for the definitions of the other couplings).}
   \label{Fig.4}
  \end{center}
\end{figure}%

\section{Work in progress}

An obvious step forward is the computation of quark bilinears in the case of other actions. A first possibility is to change the gauge action: we will take into account the Symanzick improved version. Some comments can be made for the fermionic regularization dictated by 
the Clover action, for which the knowledge of $c_{SW}$ is required. The one loop result is available, which suffices for two loops computations. Still, an accurate effective parametrization of $c_{SW}$ both for $n_f=0$ and $n_f=2$ is available, which entails the one loop result \cite{cSW}. For example, in the case $n_f=2$
\begin{equation}
	c_{SW} = \frac{1 - 0.454 \,g_0^2 - 0.175 \,g_0^4 + 0.012 \,g_0^6 + 0.045 \,g_0^8}{1 - 0.720 \,g_0^2}
	\label{c_sw}
\end{equation}
Our strategy is of course to compute the second loop coefficient for $c_{SW}$. Estimates of three loops coefficients can in the meantime be obtained by extracting from Eq.~(\ref{c_sw}) (and its quenched counterpart) an estimate for the two loops $c_{SW}$ (the rational form of Eq.~(\ref{c_sw}) is well suited to this purpose).

\end{document}